\documentclass[fp,twocolumn]{jpsj3}
\usepackage{txfonts}

\title{Splitting Fermi Surfaces and Heavy Electronic States in Non-Centrosymmetric U$_3$Ni$_3$Sn$_4$}
\author{Arvind Maurya$^1$, 
Hisatomo Harima$^2$, 
Ai Nakamura$^1$, 
Yusei Shimizu$^1$, 
Yoshiya Homma$^1$, 
DeXin Li$^1$, 
Fuminori Honda$^1$, 
Yoshiki J. Sato$^1$,
and Dai Aoki$^{1}$\thanks{aoki@imr.tohoku.ac.jp}}
\inst{
$^1$Institute for Materials Research, Tohoku University, Oarai, Ibaraki 311-1313, Japan \\
$^2$Graduate School of Science, Kobe University, Kobe 657-8501, Japan}%\\
\abst{We report the single-crystal growth of the non-centrosymmetric paramagnet U$_3$Ni$_3$Sn$_4$  by the Bridgman method 
and the Fermi surface properties detected by de Haas--van Alphen  (dHvA) experiments. 
We have also investigated single-crystal U$_3$Ni$_3$Sn$_4$ by single-crystal X-ray diffraction, 
magnetization, electrical resistivity, and heat capacity measurements. 
The angular dependence of the dHvA frequencies reveals many closed Fermi surfaces, which are nearly spherical in topology.
The experimental results are in good agreement with local density approximation (LDA) band structure calculations 
based on the $5f$-itinerant model. 
The band structure calculation predicts many Fermi surfaces, mostly with spherical shape, derived from 12 bands crossing the Fermi energy. 
To our knowledge,
the splitting of Fermi surfaces due to the non-centrosymmetric crystal in $5f$-electron systems is experimentally detected for the first time.
The temperature dependence of the dHvA amplitude reveals a large cyclotron effective mass of up to $35\,m_0$, 
indicating the heavy electronic state of U$_3$Ni$_3$Sn$_4$ due to the proximity of the quantum critical point.
From the field dependence of the dHvA amplitude, a mean free path of conduction electrons of up to 1950~$\AA$ is detected, reflecting the good quality of the grown crystal.
The small splitting energy related to the antisymmetric spin-orbit interaction is most likely due to the large cyclotron effective mass.
 }
\begin{document}
\maketitle
\section{Introduction}
Uranium compounds have been extensively studied owing to the intriguing nature of their 5$f$ electrons giving rise to anomalous effects such as unconventional superconductivity, the coexistence of ferromagnetism and superconductivity, heavy-fermion behavior, anisotropic hybridization, local and itinerant magnetism, hidden order, and multipolar interactions.~\cite{Buschow,Aoki,Fournier,Moore,Santini} 
The 5$f$ electrons in uranium compounds may behave as itinerant as well as localized nature
unlike the rather extreme cases of lanthanides (localized) and $d$\--transition metals (itinerant). 
In the last few decades, non-centrosymmetric materials have attracted condensed matter physicists owing to their potential 
to exhibit anomalous behavior, for instance, mixed pair wave unconventional superconductivity, magnetic skyrmion lattices, multipolar ordering, and chiral fermiology. 
Spin-orbit coupling (SOC) plays an important role in such compounds 
by introducing splitting in otherwise spin-degenerate bands. 
Since the strength of SOC increases with the atomic number, non-centrosymmetric materials containing heavy elements such as uranium are promising candidates for interesting phenomena introduced by the confluence of SOC and the semi-itinerant character of $5f$ electrons. 
The de Haas--van Alphen (dHvA) effect, being an indispensable tool to explore the topology of  Fermi surfaces, plays a crucial role in demystifying such an effect,
because the splitting of Fermi surfaces due to SOC in a non-centrosymmetric material is directly detected from a microscopic viewpoint.

Many U$_3$T$_3$X$_4$ non-centrosymmetric uranium compounds with the cubic crystal structure (space group: \#220, $I\bar{4}3d$, $T_d^6$) exist, 
namely U$_3$T$_3$Sn$_4$ (T~=~Pt, Ni, Au, Cu), 
U$_3$T$_3$Bi$_4$ (T~=~Rh, Ni), and
U$_3$T$_3$Sb$_4$ (T~=~Pt, Pd, Ni, Cu, Co, Rh, Ir),
possessing a rich variety of their properties: itinerant ferromagnet, metal, semiconductor, and heavy-fermion behavior~\cite{Dwight,Endstra1,Takabatake1,Takabatake2,Canfield,Klimczuk,Shlyk1,Shlyk2,Baek}. 
However, most of them are either semiconducting [U$_3$Ni$_3$Bi$_4$, U$_3$(Pd,Pt,Ni)$_3$Sb$_4$] or possess large residual scattering, 
for example, the residual resistivities, c
$\rho_0$ of U$_3$Au$_3$Sn$_4$, U$_3$Rh$_3$Bi$_4$, U$_3$Co$_3$Sb$_4$, U$_3$Cu$_3$Sn$_4$, and U$_3$Cu$_3$Sb$_4$ are 500, 150, 300, 800, and $180\,\mu\Omega\!\cdot\!{\rm cm}$, respectively~\cite{Takabatake1,Takabatake2,Canfield,Klimczuk}. 
Nevertheless, U$_3$Ni$_3$Sn$_4$ and U$_3$Pt$_3$Sn$_4$ exhibit good values of $\rho_0$ ($< 5\,\mu\Omega\!\cdot\!{\rm cm}$) and the residual resistivity ratio ($>200$) even in their polycrystalline samples~\cite{Takabatake1,Endstra1}, 
making them suitable for quantum oscillation studies. 
Moreover, their paramagnetic ground state rules out the additional complexity in the band structure arising in the magnetically ordered state due to the formation of a magnetic Brillouin zone. 
The metallic or semiconducting behavior in this homologous series has been proposed to be governed by the electron count, charge balance, hybridization strength, and the intricate band structure.

Note that none of the aforementioned compounds have been explored in terms of detailed Fermi surface properties. 
However, Inada et al.\ have studied structurally related ferromagnets U$_3$As$_4$ ($T_{\rm C}$~=~198~K) and U$_3$P$_4$ ($T_{\rm C}$~=~138~K) by dHvA and magnetoresistance measurements~\cite{Inada}. 
Most of the dHvA branches in U$_3$X$_4$ (X~=~P, As) can be qualitatively explained by the spin-polarized 5$f$-itinerant band model. 
An enhancement of the cyclotron effective mass up to 70~$m_0$ (33~$m_0$) in U$_3$As$_4$ (U$_3$P$_4$)  is in consonance with the Sommerfeld coefficient of 83 (90) mJ/(mol-U K$^2$). 
The magnetoresistance measurements confirmed the compensation of charge carriers in both metals and the existence of open orbits in U$_3$P$_4$.

U$_3$Ni$_3$Sn$_4$ is described as a heavy-fermion paramagnet located at the verge of the antiferromagnetic quantum critical point with the negative critical pressure of $-0.04\,{\rm GPa}$~\cite{Estrela,Visser,Booth,Shlyk1}. 
From low-temperature specific heat measurements, 
it is suggested that the crossover from the non-Fermi liquid to the Fermi liquid state occurs below $0.4\,{\rm K}$~\cite{Shlyk2}.
Upon applying pressure, the Fermi liquid state is recovered in a wide temperature range.~\cite{Estrela}
A photoemission study by Takabatake et al.\ indicated that hybridization of the partially filled Ni 3$d$ bands with the U 5$f$ states may lead to the delocalization of the 5$f$ electrons in U$_3$Ni$_3$Sn$_4$~\cite{Takabatake3}. 
On the other hand, Endstra et al.\ demonstrated a continuous transformation from metallic to semiconducting behavior by substituting Sn by Sb, establishing the importance of Sn 5$p$ electrons in the itineracy of U 5$f$ orbitals~\cite{Endstra2}. 
NMR studies reveal a crossover from localized to heavy-fermion character of the U 5$f$ electrons in U$_3$Ni$_3$Sn$_4$~\cite{Kojima} as the temperature is lowered. 
Bremsstrahlung isochromat spectroscopy suggests that the hybridization of U $5f$  with Ni $3d$ and Sn $5p$ states as well as the Coulomb repulsion between the 5$f$-electrons is non-negligible~\cite{Ejima}. 
These findings suggest that the intricate correlations of electron orbitals drive U$_3$Ni$_3$Sn$_4$ to a heavy-fermion paramagnetic metal.

In order to clarify the electronic state of U$_3$Ni$_3$Sn$_4$, 
we carried out dHvA experiments
in addition to magnetic susceptibility, magnetization, resistivity, and heat capacity measurements
using a high-quality single crystal grown by the Bridgman method with a large residual resistivity ratio of ${\rm RRR}\sim 300$.
The detected Fermi surfaces are nearly spherical in topology with large cyclotron effective masses,
indicating a heavy electronic state.
The results are in reasonably good agreement with local density approximation (LDA) band calculations based on the 5$f$-itinerant model.
The splitting of Fermi surfaces is detected because of the SOC owing to the lack of inversion symmetry in the crystal structure.
The small splitting energies are probably due to the heavy cyclotron effective masses.

% ================================================================================
\section{Experimental Procedure}
We have utilized the Bridgman method to grow U$_3$Ni$_3$Sn$_4$ single crystals. 
A polycrystalline ingot of U$_3$Ni$_3$Sn$_4$ with a stoichiometric amount of each element was prepared in a tetra arc furnace before sealing it inside a tungsten Bridgman crucible in argon atmosphere. 
The sealed tungsten crucible was placed in a vertical tube furnace with a precalibrated temperature and thermal gradient profile. 
Argon gas was continuously flowed to avoid the oxidation of the crucible. 
The temperature inside the furnace was raised to 1350~$^{\circ}$C in 12 h, 
and maintained at this temperature for another 12 h to ensure complete melting of the charge inside the crucible. 
After that, it was cooled  to 900~$^{\circ}$C at a rate of 4~$^{\circ}$C/h 
then to 600~$^{\circ}{\rm C}$ at a faster rate of 30~$^{\circ}$C/h and finally to room temperature at 60~$^{\circ}$C/h. 
 
Clear Laue diffraction spots indicated the good quality of the grown crystal. 
However, it also turned out that the lump in the crucible after crystal growth contained several domains. 
Monodomain crystals were isolated by means of Laue photographs and a spark cutter. 
The maximum grain size was approximately $3\times 2\times 1\,{\rm mm^3}$. 
Laue photographs were taken from all sides of the isolated crystal to ensure that it was a monodomain. 
The phase purity was confirmed by powder X-ray diffraction. 
Moreover, single-crystal X-ray diffraction was performed to refine the lattice parameters and atomic positions, 
which were subsequently used as the input in the band structure calculations.

The electrical resistivity in the temperature range $300$--$0.2\,{\rm K}$ was measured by a laboratory-built adiabatic demagnetization refrigerator (ADR) coupled with a Quantum Design physical property measurement system (QD PPMS). 
Heat capacity and magnetization measurements were carried out in the QD PPMS and a QD magnetic property measurement system (MPMS), respectively. 
The dHvA measurements at low temperatures down to 35~mK and at high fields up to 147~kOe were carried out in a top-loading dilution refrigerator using a standard field modulation technique.

% ================================================================================
\section{Results and Discussion}
\begin{figure}[!]
\begin{center}
\includegraphics[width=.4\textwidth]{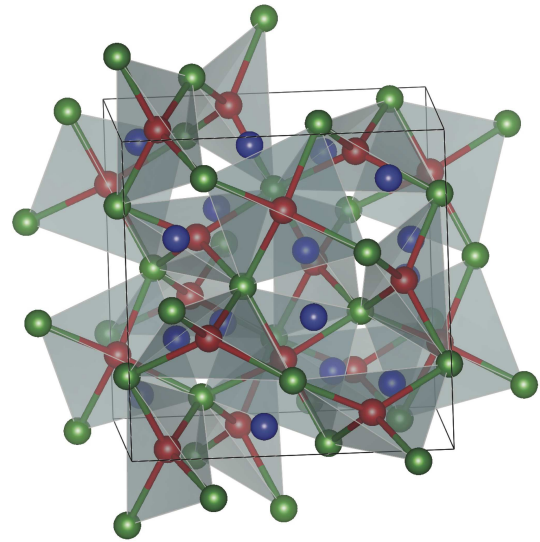}
\caption{\label{Fig1}(Color online) Crystal structure of U$_3$Ni$_3$Sn$_4$. Corner-sharing tetrahedra represent the crystallographically rigid network of U (red) and Ni (green) atoms placed at special crystallographic sites, while Sn atoms (blue) occupy a general position.}
\end{center}
\end{figure}

U$_3$Ni$_3$Sn$_4$ crystallizes in an Y$_3$Au$_3$Sb$_4$-type non-centrosymmetric body-centered cubic structure, which is a ``filled up'' version of the Th$_3$P$_4$\--type structure~\cite{Dwight}, as shown in Fig.~\ref{Fig1}.
In order to determine the crystallographic parameters of U$_3$Ni$_3$Sn$_4$,
we performed single-crystal X-ray diffraction at room temperature
and analyzed the data by least-squares refinement.
The results are summarized in Tabled~\ref{Table1} and \ref{Table2}.
The lattice parameter was determined to be $a$~=~9.3575(4)~$\AA$.
In U$_3$Ni$_3$Sn$_4$ crystal, U and Ni atoms occupy the Wyckoff sites 12$a$ and 12$b$, respectively, of space group $I\bar{4}3d$~(\#220, $T_d^6$) possessing fourfold rotational symmetry. 
Note that these sites have fixed special crystallographic positions and form a rigid network of sites 
coordinated tetrahedrally with respect to each other. 
Sn atoms are placed at the site labeled as 16$c$ having threefold axial symmetry along $\langle111\rangle$. 
The position of Sn atoms is specified by a single variable identical for $x$, $y$, and $z$ atomic coordinates.
The fractional coordinate for the Sn position is found to be 0.0820(2).
The obtained lattice parameter and fractional coordinates are close to the previously reported values~\cite{Dwight,Shlyk1}. 
Although the unit cell is large, the smallest distance between U atoms, $d_{\rm U-U}$, is merely 4.377~\AA. 
The rather small values of $R1$ and $wR2$ give credence to the correctness of our refined parameters over those in previous reports~\cite{Dwight,Shlyk1}.

%
%*********************FIGURE 1 **********************************
%****************************************************************
%
%
%****************************Table1******************************
\begin{table}[!]
\caption{\label{Table1} Crystal data and structure refinement parameters of U$_3$Ni$_3$Sn$_4$ single crystal.}
\begin{center}
\begin{tabular}{cc}
\hline
Parameter 					& 	Value\\ \hline
Structure prototype 		&	Y$_3$Au$_3$Sb$_4$  \\
Crystal system 				& 	cubic  \\
Space group					&	 $I\bar{4}3d$, \#220, $T_d^6$\\
$Z$							& 	 4\\
Lattice parameter [\AA]		& 	$a$~=~9.3575(4)\\
Radiation					& 	MoK$_\alpha$ ($\lambda$~=~0.71073~\AA) \\
						& 	(graphite monochromated)\\
Temperature [$^\circ$C]			& 		23.0\\
$\mu$(MoK$_\alpha$) [mm$^{-1}$]& 	77.646\\
Calculated density [g cm$^{-3}$]	& 	11.064\\
Scan mode					& 	$\omega-2\theta$\\
Measured reflections		& 	1087\\
Independent reflections		& 	253\\
Refinement					&	  Full-matrix least-squares on $F^2$\\
2$\theta_{\rm max}$					&	 66.1$^\circ$\\
Refined parameters			& 	9\\
Extinction coefficient		& 	0.0365\\
$R1$							& 	0.027\\
$wR2$ (All reflections)		& 	0.0903\\
Goodness-of-fit				& 	1.074\\
\hline
\end{tabular}
\end{center}
\end{table}
%****************************************************************
%
%****************************Table2******************************
\begin{table}[!]
\caption{\label{Table2} Fractional atomic coordinates of U$_3$Ni$_3$Sn$_4$ as determined from single-crystal X-ray diffraction data.}
\begin{center}
\begin{tabular}{c@{\,}c@{\,}c@{\,}c@{\,}c@{\,}c}
\hline
Atoms		&Wyckoff site	&Site symmetry	&$x$		&$y$		& $z$\\ \hline
U			&12$a$			&$\bar{4}$		&3/8		&0		&1/4\\
Ni			&12$b$			&$\bar{4}$		&7/8		&0		&1/4\\
Sn			&16$c$			&3				&0.0820(2)	&0.0820(2)&0.0820(2)\\
\hline
\end{tabular}
\end{center}
\end{table}
%****************************************************************
%
%*********************FIGURE 2 **********************************
\begin{figure*}[t]
\begin{center}
\includegraphics[width=0.8\textwidth]{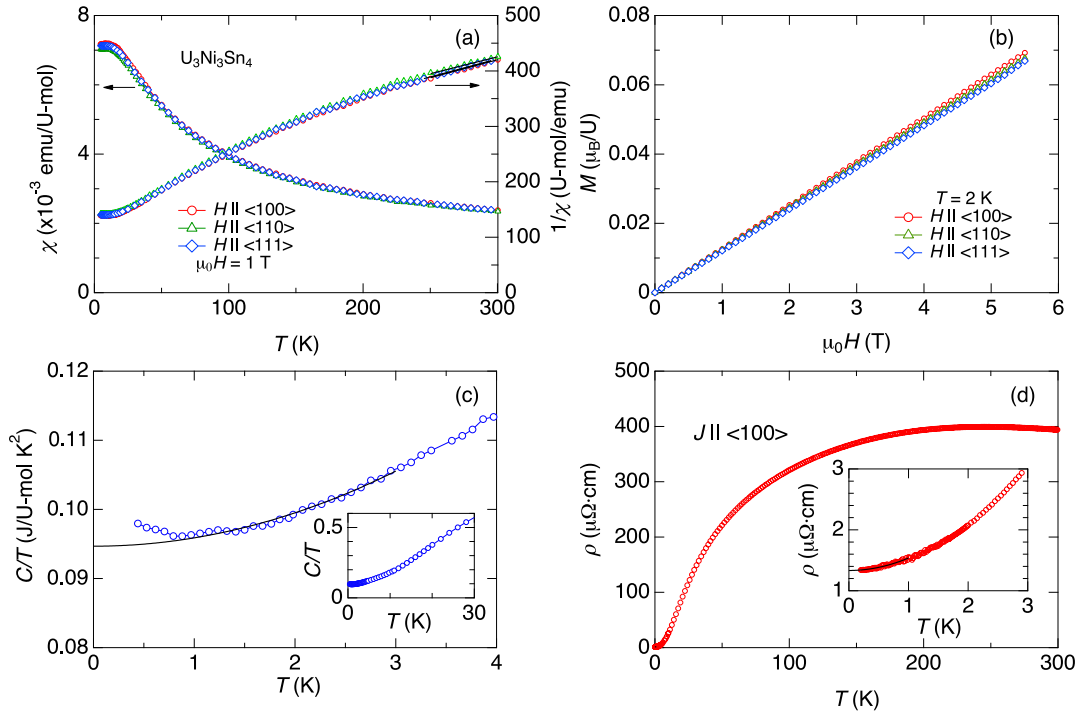}
\caption{\label{Fig2}(Color online) Physical properties as a function of temperature of single-crystal U$_3$Ni$_3$Sn$_4$: (a) magnetic susceptibility (left axis) and inverse susceptibility (right axis), (b) isothermal magnetization at 2~K, (c) specific heat, and (d) electrical resistivity. Solid black lines in panels (a), (c), (d) represent the fitted curves as discussed in the text.}
\end{center}
\end{figure*}
%****************************************************************
%

Figures~\ref{Fig2}(a) and \ref{Fig2}(b) show the isotropy in the magnetic susceptibility measured in a field of 1~T field and the isothermal magnetization at 2~K along crystallographic directions $\langle100\rangle$, $\langle110\rangle$, and $\langle111\rangle$, respectively. 
This is not unexpected considering the isotropically dense unit cell. 
The magnetization appears to be linear and attains a value of only $\sim 0.07\,\mu_{\rm B}/{\rm U}$ at 5.5~T.  
The susceptibility increases with decreasing temperature and suturates below 15~K to a value of $\chi_0 \approx 7\times 10^{-3}\,{\rm emu/U\mbox{-}mol}$. 
This behavior is reminiscent of the crossover from Curie to Pauli susceptibility below the Kondo temperature in Kondo lattice compounds. 
The inverse susceptibility ($\chi^{-1}$) is a nonlinear function of temperature, although the Curie-Weiss law, $\chi^{-1}=(T-\theta_{\rm P})/C$ can be fitted in the temperature range 250--300~K, 
giving effective paramagnetic moment ($\mu_{\rm eff}$) values of 3.51, 3.63, and 3.50 $\mu_{\rm B}/\rm U$ along the $\langle 100\rangle$, $\langle 110\rangle$, and $\langle 111\rangle$ directions, respectively.   
The $\mu_{\rm eff}$ values are comparable to those for the  U$^{3+}$~($5f^3$) or U$^{4+}$~($5f^2$) configuration of the U ion. 

Figure~\ref{Fig2}(c) show the specific heat as a function of temperature [$C(T)$] for U$_3$Ni$_3$Sn$_4$ at temperatures down to $0.4\,{\rm K}$. 
Fitting the equation $C/T=\gamma+\beta T^2$ to the low-temperature data 
in the range $1$ to $3\,{\rm K}$
gives the electronic specific heat coefficient (Sommerfeld coefficient) of $\gamma = 95\,{\rm mJ/(U\mbox{-}mol\, K^2)}$, 
which is close to the values for a polycrystalline sample~\cite{Takabatake1} and a single-crystal sample~\cite{Shlyk2}.
The small upturn below $1\,{\rm K}$ is also consistent with the previous results in a single crystal, 
although the $C/T$ value here is slightly lower than the previous ones.
This upturn is probably related to the crossover from the non-Fermi liquid to Fermi liquid regime around $0.4\,{\rm K}$ together with the nuclear contribution to the specific heat, 
as mentioned in Ref.~\citen{Shlyk2}.

Electrical resistivity as a function of the temperature, $\rho(T)$, of U$_3$Ni$_3$Sn$_4$ for $J\parallel \langle 100\rangle$ is shown in Fig.~\ref{Fig2}(d). 
The resistivity follows the $T^2$ dependence below $1\,{\rm K}$,
and a least-squares fitting of the expression $\rho$($T$)~=~$\rho_0+AT^2$ gives residual resistivity 
$\rho_0=1.3\,\mu\Omega\!\cdot\!{\rm cm}$ and 
$A=0.20\,\mu\Omega\!\cdot\!{\rm cm\,K^{-2}}$.
The large value of coefficient $A$ results a Kadowaki-Woods ratio ($A/\gamma^2$) of 
$2\times 10^{-5}\,\mu\Omega\!\cdot\!{\rm cm (mJ/K\,U\mbox{-}mol)^2}$,  
which is slightly higher but not very far from the universal value 
[$1\times 10^{-5}\,\mu\Omega\!\cdot\!{\rm cm (mJ/K\,mol)^2}$]. 
Also the Wilson ratio $R_{\rm W}$, defined as 
$R_{\rm W}=\pi^2k_{\rm B}^2\chi_0/(\gamma\mu_{\rm eff}^2)$,
for U$_3$Ni$_3$Sn$_4$ is 1.3, close to the free-electron value of unity.  
$\rho$($T$) at 300~K ($394\,\mu\Omega\!\cdot\!{\rm cm}$) in conjunction with the value of $\rho_0$ gives a residual resistivity ratio ($RRR=\rho_{\rm RT}/\rho_0$)  of 296, 
indicating the high quality of the grown U$_3$Ni$_3$Sn$_4$ crystal. 
The broad hump in $\rho(T)$ around 250~K may be attributed to the hybridization of U 5$f$ electrons with the conduction electrons.

\begin{figure}[t]
\begin{center}
\includegraphics[width=0.45\textwidth]{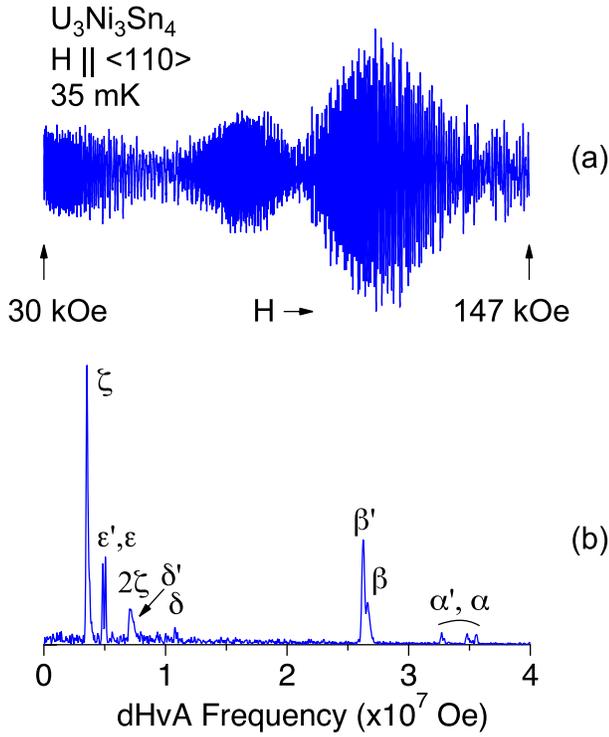}
\caption{\label{dhva_1}(Color online) dHvA oscillations and corresponding FFT spectrum at 35~mK along $\langle 110\rangle$ in  U$_3$Ni$_3$Sn$_4$.}
\end{center}
\end{figure}

% dHvA
The angular dependence of dHvA oscillations was measured from $\langle110\rangle$ to $\langle100\rangle$ and from $\langle100\rangle$ to $\langle111\rangle$ then $\langle110\rangle$ by rotating the crystal in the $\lbrace001\rbrace$ and  $\lbrace110\rbrace$ planes, respectively. 
The dHvA signal starts to appear from a small field of $\sim 1.5\,{\rm T}$,
indicating the high quality of the present sample.
Figure~\ref{dhva_1} shows the typical dHvA oscillations in the field range 3--14.7~T and the corresponding FFT spectra for the magnetic field parallel to the $\langle110\rangle$ crystallographic directions at 35~mK.  
dHvA oscillations of frequency components of up to $\sim 4\times 10^7\,{\rm Oe}$ were observed, which can be categorized 
into branches $\alpha$, $\alpha^\prime$, $\beta$, $\beta^\prime$, $\delta$, $\delta^\prime$, $\varepsilon$, $\varepsilon^\prime$, and $\zeta$ as fundamental dHvA branches.
%
%*********************FIGURE 3 **********************************
%****************************************************************
%

The splitting of dHvA branches except for branch $\zeta$ is observed in Fig.~\ref{dhva_1}.
In general, the splitting of the dHvA frequency is often accounted for 
(a) similar extremal orbits of the same (often corrugated) Fermi surfaces, 
(b) spin-split Fermi surfaces with a small energy difference,
(c) magnetic breakdown between two Fermi surfaces closely separated in energy, or
(d) bicrystals with slightly misaligned grains. 
However, we observed a similar spectrum for a single-crystal obtained from a different batch. 
This leads us to conclude that the splitting in this case is primarily due to antisymmetric-SOC-induced lifting of the twofold spin degeneracy and the magnetic breakdown effect. 
The phenomena of magnetic breakdown occur when the energy gap between Fermi surfaces is small and electrons in the cyclotron orbits possess enough energy to tunnel or ``break through" the gap at high magnetic fields. 
This may modify the spectra of dHvA frequencies, which sometimes show forbidden orbits with extremal area larger than the Brillouin zone~\cite{Shoenberg}. 
Blount derived the criterion for magnetic breakdown as $\hbar\omega_{\rm C} \gtrsim \varepsilon_{\rm g}^2/\varepsilon_{\rm F}$, 
where $\omega_{\rm C}$ is the cyclotron frequency, 
$\varepsilon_{\rm g}$ is the energy gap, 
and $\varepsilon_{\rm F}$ is the Fermi energy~\cite{Blount_MB}. 
This condition can be easily realized for spin-split Fermi surfaces, particularly in high-symmetry cases. 
One recent example is TaSi$_2$, in which two additional cyclotron orbits were detected by the magnetic breakdown originating from lack of an inversion center in its crystal structure~\cite{Nakamura_TaSi2}.
 
%
%*********************FIGURE 4 **********************************
\begin{figure*}[t]
\begin{center}
\includegraphics[width=0.95\textwidth]{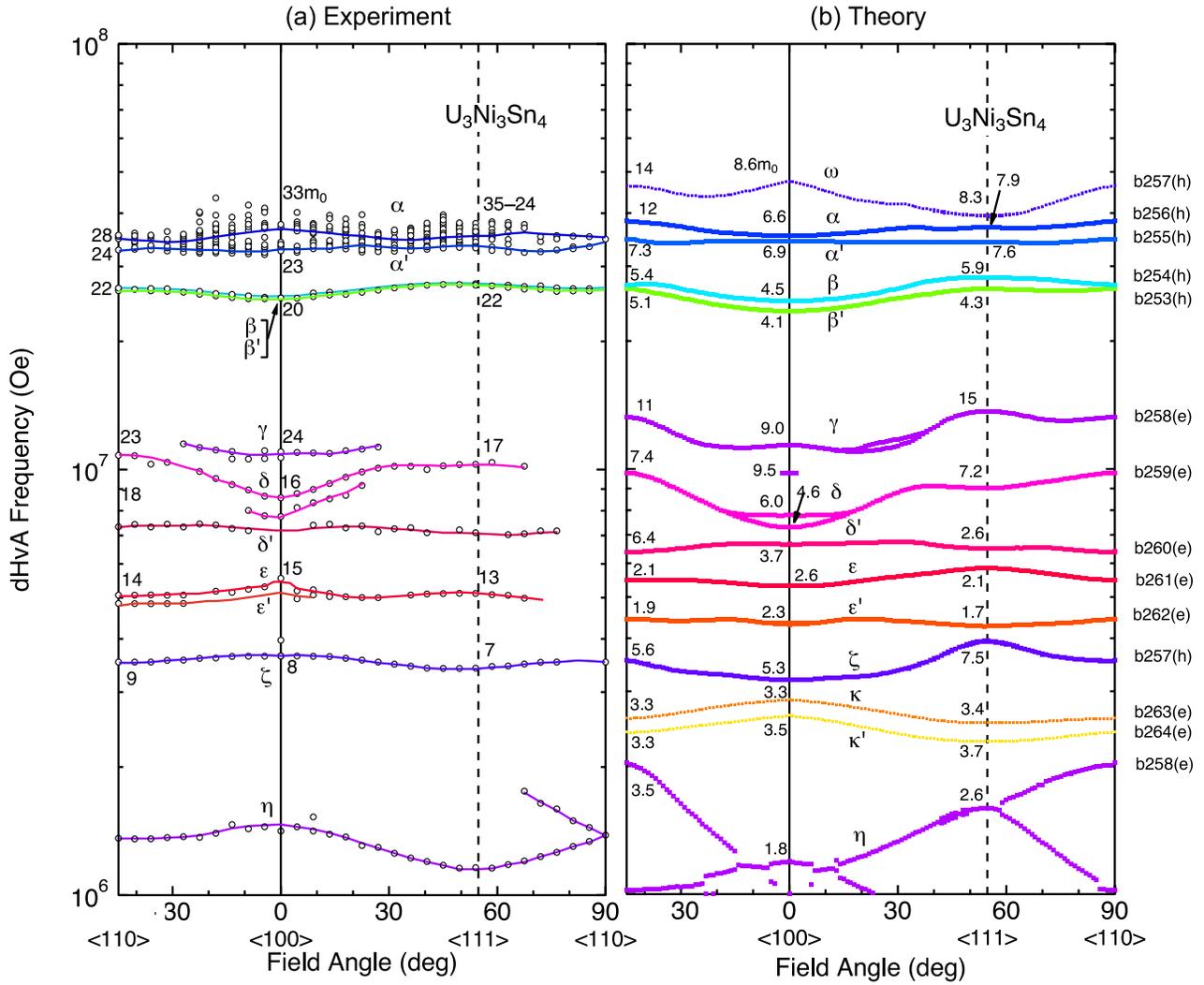}
\caption{(Color online) Angular dependence of the dHvA frequencies in U$_3$Ni$_3$Sn$_4$ obtained by (a) experiment and (b) theory based on the LDA band calculations. Cyclotron effective masses (a) and band masses (b) are also shown as numbers for $H\parallel \langle 100\rangle$, $\langle 110\rangle$, and $\langle 111\rangle$.
The thick solid lines in panel (b) correspond to the experimentally detected dHvA branches.
The annotations at the right side in panel (b) indicate the band number and the hole/electron Fermi surface. 
For instance,
``b256(h)'' corresponding to branch $\alpha$ indicates the band 256th hole Fermi suface.}
\label{fig:AngDep}
\end{center}
\end{figure*}

%****************************************************************
%
Figure~\ref{fig:AngDep}(a) shows the dHvA frequencies as a function of sample orientation with respect to the magnetic field.  
Ten different dHvA branches, namely $\alpha$, $\alpha^\prime$, $\beta$, $\beta^\prime$, $\gamma$, $\delta$, $\delta^\prime$, $\varepsilon$, $\zeta$, and $\eta$, were detected. 
Most of the frequencies are nearly constant over a wide range of angles, which is expected for spherical Fermi surfaces. 
Under this scenario, branches 
$\alpha$, $\alpha^\prime$, $\beta$, $\beta^\prime$, $\gamma$, $\delta$, $\delta^\prime$, $\varepsilon$, $\zeta$, and $\eta$ occupy 25\%, 22\%, 16\%, 16\%, 4.2\%, 3.6\%, 2.3\%, 1.4\%, 0.77\%, and 0.17\%  
of volume of the Brillouin zone, respectively. 

%
%****************************Table4******************************
\begin{table}[h]
\newcommand{\N}{\phantom{0}}
\caption{Experimental dHvA frequency $F$, cyclotron effective mass $m_{\rm c}^\ast$, calculated dHvA frequency $F_{\rm b}$ and band mass $m_{\rm b}$ in U$_3$Ni$_3$Sn$_4$.}
\label{tab}
\begin{center}
\begin{tabular}{ccccc}

\hline
				& \multicolumn{2}{c}{Experiment}				& \multicolumn{2}{c}{Theory}	\\
Branch 			& $F$ 					 & $m_{\rm c}^\ast$ 	&  $F_{\rm b}$			& $m_{\rm b}$\\ 
			    &($\times10^6\,{\rm Oe}$)& ($m_0$)			 	&($\times10^6\,{\rm Oe}$)& $(m_0)$    \\
\hline
%------------
$H\parallel \langle100\rangle$\\
$\omega$		&						& 						& 47.5					& 8.6		\\
$\alpha$		& 37.6					& 33					& 35.4					& 6.6		\\
$\alpha^\prime$	& 32.6					& 23					& 34.4					& 6.9		\\
$\beta$			& 25.3					& 20					& 24.8					& 4.5		\\
$\beta^\prime$	& 						&						& 23.5					& 4.1		\\
$\gamma$		& 11.0					& 24					& 11.4					& 9.0		\\
				&						&						& 9.78					& 9.5		\\
$\delta$		& 8.58					& 16					& 7.78					& 6.0		\\
			 	& 7.75					&						& 7.29					& 4.6		\\
$\delta^\prime$ &						&						& 6.65					& 3.7		\\
$\varepsilon$	& 5.55					& 15					& 5.32					& 2.6		\\
$\varepsilon^\prime$ &					&						& 4.34					& 2.3		\\
$\zeta$			& 3.65					& 8						& 3.20					& 5.3		\\
$\kappa$		&						&						& 2.87					& 3.3		\\
$\kappa^\prime$ & 						&						& 2.63					& 3.5		\\
$\eta$			& 1.41					&						& 1.19					& 1.8		\\ 
\hline
%------------
$H\parallel \langle110\rangle$\\
$\omega$		&						& 						& 46.4					& 14		\\
$\alpha$		& 34.8					& 28					& 38.3					& 12		\\
$\alpha^\prime$	& 32.8					& 24					& 34.7					& 7.3		\\
$\beta$			& 26.7					& 22					& 27.1					& 5.4		\\
$\beta^\prime$	& 26.3					& 						& 26.6					& 5.1		\\
$\gamma$		& 						& 						& 13.2					& 11		\\
$\delta$		& 10.8					& 23					& 9.79					& 7.4		\\
$\delta^\prime$ & 7.33					& 18					& 6.39					& 6.4		\\
$\varepsilon$	& 5.07					& 14					& 5.48					& 2.1		\\
$\varepsilon^\prime$ & 4.84				&						& 4.41					& 1.9		\\
$\zeta$			& 3.53					& 9						& 3.55					& 5.6		\\
$\kappa$		& 						&						& 2.59					& 3.3		\\
$\kappa^\prime$ & 						&						& 2.40					& 3.3		\\
$\eta$			& 1.36					&						& 2.04					& 3.5		\\ 
\hline
%------------
$H\parallel \langle111\rangle$\\
$\omega$		&						& 						& 39.6					& 8.3		\\
$\alpha$		& 38.3--33.7			& 35--24				& 37.0					& 7.9		\\
$\alpha^\prime$	&  						&  						& 34.1					& 7.6		\\
$\beta$			& 27.3					& 22					& 28.2					& 5.9		\\
$\beta^\prime$	& 27.1					& 						& 26.5					& 4.3		\\
$\gamma$		& 						& 						& 13.7					& 15.1		\\
$\delta$		& 10.3					& 17					& 9.01					& 7.2		\\
$\delta^\prime$ & 7.10					& 						& 6.51					& 2.6		\\
$\varepsilon$	& 5.11					& 13					& 5.85					& 2.1		\\
$\varepsilon^\prime$ & 					&						& 4.27					& 1.7		\\
$\zeta$			& 3.41					& 7						& 3.94					& 7.5		\\
$\kappa$		& 						&						& 2.53					& 3.4		\\
$\kappa^\prime$ & 						&						& 2.29					& 3.7		\\
$\eta$			& 1.16					&						& 1.59					& 2.6		\\ 
\hline

\end{tabular}
\end{center}
\end{table}
%****************************************************************
%
In order to clarify the Fermi surfaces in U$_3$Ni$_3$Sn$_4$, 
we performed energy band structure calculations using the KANSAI code 
based on an FLAPW (full potential linear augmented plane wave) method with an LDA for the $5f$-itinerant model. 
In the band calculations, the scalar relativistic effect was taken into account for all the electrons.
The SOC was included as the second variational procedure for valence electrons. 
In the calculations, we used the structural parameters listed in Tables~\ref{Table1} and \ref{Table2}. 
Core electrons (Rn-core minus $6s^26p^6$ electrons for U, Ar-core for Ni, Kr-core for Sn) were calculated inside muffin-tin spheres in each self-consistent step. 
$6s^26p^6$ electrons on U and $4d^{10}$ electrons on Sn were calculated as valence electrons by using a second energy window. 
The LAPW basis functions were truncated at $|k+G_i| \le 9.20 \times 2\pi/a$, corresponding to 1601 LAPW functions. 
The sampling points were uniformly distributed to 506 k-points in the irreducible 1/24th of the Brillouin zone, which were divided by (20, 20, 20).

\begin{figure}[tbhp]
\begin{center}
\includegraphics[width=0.45\textwidth]{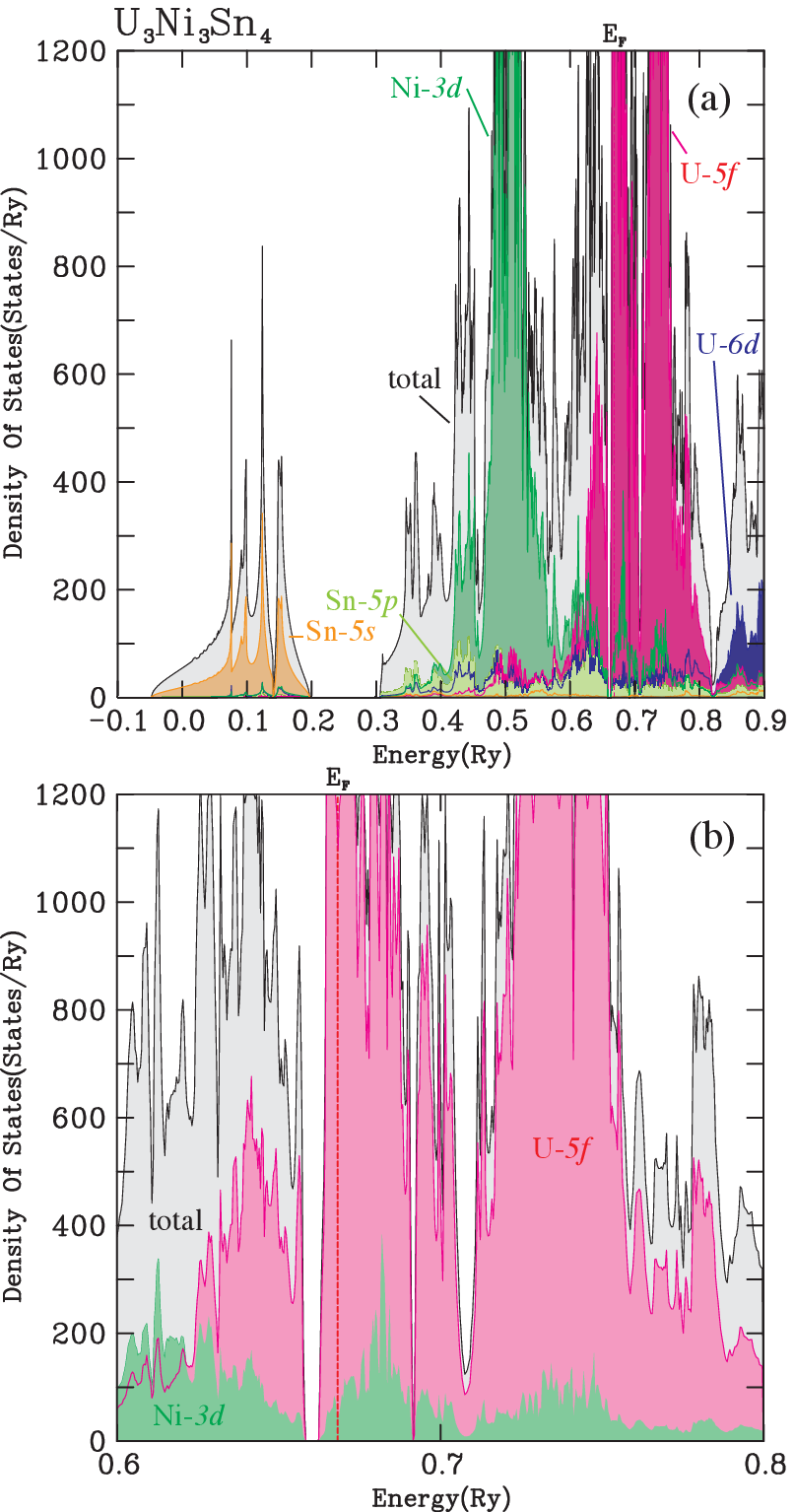}
\caption{(Color online) 
Calculated density of states for U$_3$Ni$_3$Sn$_4$ (a) in whole energy region and (b) near the Fermi level. 
The partial components inside the muffin-tin spheres are also shown.}
\label{fig:dos}
\end{center}
\end{figure}

Figure~\ref{fig:dos} shows the calculated density of states (DOS) in U$_3$Ni$_3$Sn$_4$.
The large DOS at the Fermi level is mainly due to the $5f$ electrons of U. 

\begin{figure}[tbhp]
\begin{center}
\includegraphics[width=0.45\textwidth]{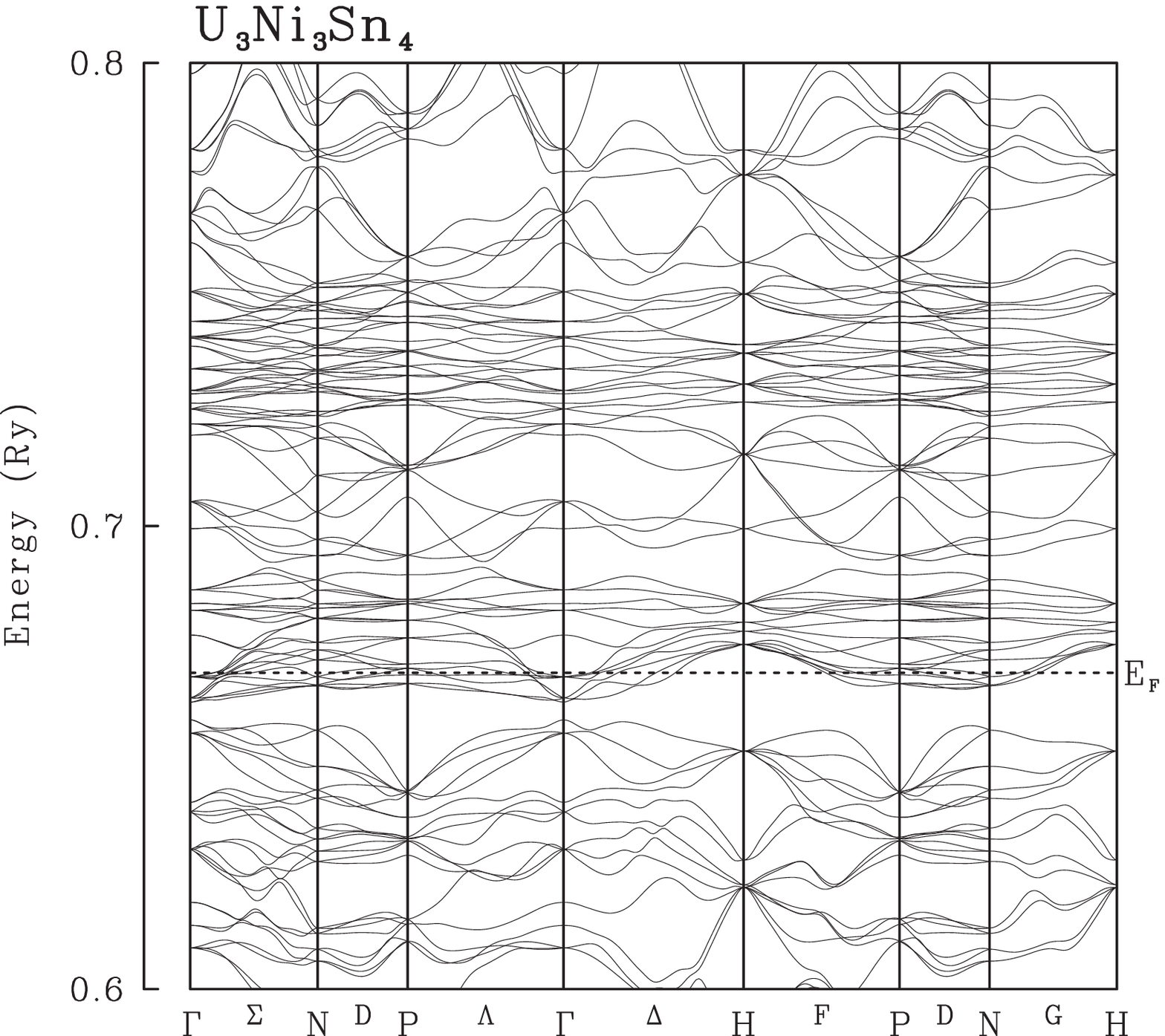}
\caption{Calculated band structure near the Fermi level. 
The $J=5/2$ bands of $5f$ electrons in U are located in the vicinity of the Fermi level. 
The $J=7/2$ bands are around $0.74\,{\rm Ry}$. See also Fig.~\ref{fig:dos}(b).}
\label{fig:bandW}
\end{center}
\end{figure}

Figure~\ref{fig:bandW} shows the band structure in U$_3$Ni$_3$Sn$_4$.
It was found that many bands in the $J=5/2$ manifold cross the Fermi level. 
Note that the bands near the Fermi level are eightfold degenerated at the H point. 
In general, fourfold degeneracy can be found in high-symmetry points in the symmorphic space group of cubic systems. 
However, U$_3$Ni$_3$Sn$_4$ belongs to the non-symmorphic space group, 
therefore the degeneracy of many of its electronic states is doubled 
at the Brillouin zone boundary. 
Such high degeneracy might be the reason why many Fermi surfaces appear in U$_3$Ni$_3$Sn$_4$.

%********************* **********************************
%****************************************************************
%********************* **********************************
%****************************************************************

\begin{figure*}[t]
\begin{center}
\includegraphics[width=0.95\textwidth]{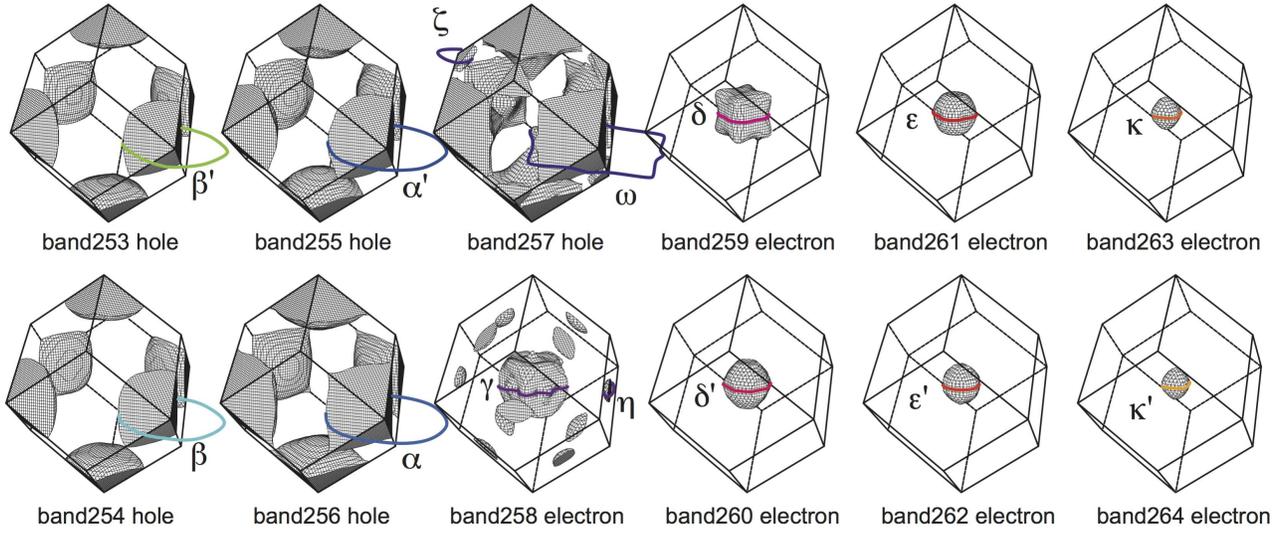}
\caption{(Color online) Calculated Fermi surfaces in the bcc Brillouin zone of U$_3$Ni$_3$Sn$_4$. }
\label{fig:FS}
\end{center}
\end{figure*}

The calculated angular dependences of U$_3$Ni$_3$Sn$_4$ based on the LDA method are compared with the fundamental frequencies observed experimentally in Fig.~\ref{fig:AngDep}(b). 
The theoretical angular dependences of dHvA frequencies are in reasonably good agreement with those obtained by experiments.
Figure~\ref{fig:FS} shows the calculated Fermi surfaces in U$_3$Ni$_3$Sn$_4$.
Twelve bands cross the Fermi energy and form the Fermi surfaces.
All of them are closed Fermi surfaces with a nearly spherical shape.

One can recognize that the Fermi surfaces are paired. 
For instance, the band 253rd hole Fermi surface is paired with the band 254th hole Fermi surface,
meaning that the original band with spin degeneracy splits into two bands due to the antisymmetric spin-orbit interaction in the non-centrosymmetric crystal structure.

The comparison between experiments and theory is summarized as follows.
\begin{enumerate}
\item
Branches $\alpha$ and $\alpha^\prime$ correspond to the band 256th and 255th hole Fermi surfaces centered at the H point, respectively. 
Many dHvA frequencies were detected near branches $\alpha$ and $\alpha^\prime$ in Fig.~\ref{fig:AngDep}(a).
This is most likely due to the magnetic breakdown between $\alpha$ and $\alpha^\prime$.

\item
Branches $\beta$ and $\beta^\prime$ correspond to the band 254th and 253rd hole Fermi surfaces centered at the H point, respectively.

\item
Branch $\gamma$ originates from the band 258th electron Fermi surface centered at the $\Gamma$ point. 
The dHvA frequency was detected only for the field close to the $\langle 100\rangle$ direction, probably due to the curvature factor.

\item
Branches $\delta$ and $\delta^\prime$ are due to the band 259th and 260th electron Fermi surfaces centered at the $\Gamma$ point. 
Since the Fermi surface of the band 259th is slightly corrugated compared with the spherical shape, the dHvA frequency splits near $H\parallel \langle 100\rangle$.

\item
Branches $\varepsilon$ and $\varepsilon^\prime$ are ascribed to the band 261st and 262nd electron Fermi surfaces centered at the $\Gamma$ point, respectively.

\item
Branch $\zeta$ is due to the band 257th hole Fermi surface centered at the P point.

\item
Branch $\eta$ originates from the small pocket Fermi surface centered at the N point in the 258th band.

\item
Branch $\omega$ originates from the largest hole Fermi surface centered at the H point in the 257th band. This Fermi surface was not detected experimentally, probably due to the large cyclotron effective mass.

\item
Branches $\kappa$ and $\kappa^\prime$ originates from the band 263rd and 264th electron Fermi surfaces centered at the $\Gamma$ point, respectively. These Fermi surfaces were not experimentally detected.

\end{enumerate}
%
%********************* **********************************
%****************************************************************

Note that
in Fig.~\ref{fig:bandW}, the splitting bands and their pairs are not very clear,
and they are rather complicated, crossing each other.
In reality, the manifold of $5f$ bands in U splits and forms 12 bands crossing the Fermi level
because of the antisymmetric SOC in the non-centrosymmetric structure.

In order to determine the cyclotron effective mass,
we measured the temperature dependence of the dHvA amplitude in the range 35--200~mK along $\langle100\rangle$, $\langle110\rangle$, and $\langle111\rangle$. 
Rapid suppression of the dHvA amplitude revealed the enhanced effective mass $m_{\rm c}^\ast$. 
We deduced values of $m_{\rm c}^\ast$ ranging from 7 to 35~$m_0$ in U$_3$Ni$_3$Sn$_4$, where $m_0$ is the rest mass of an electron. 
The large cyclotron effective mass indicates that U$_3$Ni$_3$Sn$_4$ is a heavy-fermion compound.
These values are compared with the band mass in Table~\ref{tab}.
The cyclotron effective masses are larger than the corresponding band masses with the ratio $m_{\rm c}^\ast/m_{\rm b}\sim 1\mbox{--}6$.
For example, the cyclotron effective mass of branch $\alpha$, which has the largest Fermi surface detected experimentally,
is five times larger than the band mass for $H\parallel \langle 100\rangle$.
The electronic specific heat coefficient [$\gamma\sim 95\,{\rm mJ/(K^2 U\mbox{-}mol})$] is five times larger than
that obtained from the band calculation [$\gamma_{\rm b}\sim 19.6\,{\rm mJ/(K^2 U\mbox{-}mol})$];
thus, the mass enhancement for the dHvA branch is roughly consistent with the enhancement of the $\gamma$-value.

For a spherical  Fermi surface, the electronic specific heat coefficient ($\gamma$-value)
is described as
$\gamma=\frac{\pi^2}{3}k_{\rm B}^2 D(\varepsilon_{\rm F})
=(k_{\rm B}^2 V/6\hbar^2) m_{\rm c}^\ast k_{\rm F}$,
where $D(\varepsilon_{\rm F})$ is the DOS at the Fermi level, 
$k_{\rm F}$ is half of the caliper dimension of the Fermi surface, 
and $V$ is the molar volume of U$_3$Ni$_3$Sn$_4$. 
Note that the spin degeneracy is not taken into account here because of the lack of inversion symmetry in the crystal structure.

Assuming spherical Fermi surfaces, the values of $\gamma$ derived for various dHvA frequencies vary between 1 and 13 mJ/(U-mol K$^2$). 
The sums of the values of $\gamma$ in $\langle100\rangle$, $\langle110\rangle$, and $\langle111\rangle$ amount to 36, 35, and 33~mJ/(U-mol K$^2$), respectively, which are smaller than the experimentally determined value of 95~mJ/(U-mol K$^2$). 
For $H\parallel \langle 100\rangle$, 
the missing Fermi surface $\omega$ will give $17\,{\rm mJ/K^2 U\mbox{-}mol}$, assuming the same mass enhancement of branch $\alpha$,  that is, $m_{\rm c}^\ast/m_{\rm b}\sim 5$. 
The other undetected Fermi surfaces, namely $\beta^\prime$, $\delta^\prime$, $\varepsilon^\prime$, $\kappa$, $\kappa^\prime$, and $\eta$, will give approximately $15\,{\rm mJ/K^2U\mbox{-}mol}$.
Therefore the $\gamma$-value amounts to $\sim 70\,{\rm mJ/K^2 U\mbox{-}mol}$ in total,
which is roughly consistent with the $\gamma$-value obtained by specific heat measurements.

\begin{figure}[h]
\begin{center}
\includegraphics[width=.45\textwidth]{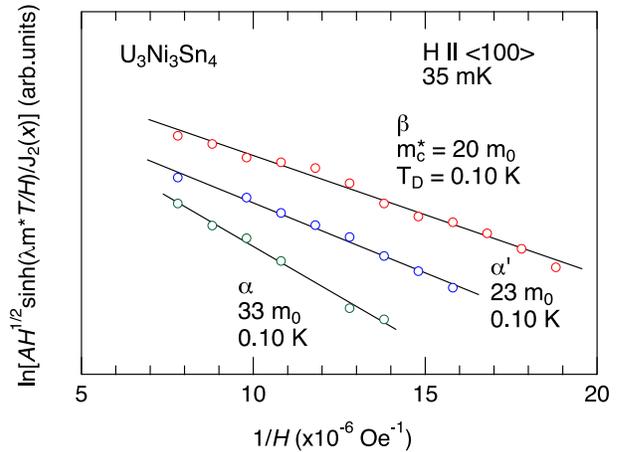}
\caption{\label{Dingle}(Color online) Dingle plot of U$_3$Ni$_3$Sn$_4$ for $H~\parallel$~$\langle100\rangle$ at $35\,{\rm mK}$.}
\end{center}
\end{figure}

According to the Lifshitz-Kosevich formula, the field dependence of the amplitude of dHvA oscillations at a constant temperature $T$ is given by
\begin{equation*}
\ln\left[AH^{1/2}\sinh\left(\lambda m_{\rm c}^*T/H\right)/J_2(x)\right]=-\lambda m_{\rm c}^* T_{\rm D}.\frac{1}{H}+ \mbox{const.},
\end{equation*}
where $A$ is the dHvA amplitude, $T_{\rm D}$ is the Dingle temperature, $\lambda$ is defined as $\lambda = 2\pi^2ck_{\rm B}/e\hbar$, and $J_2(x)$ is the Bessel function due to the field modulation technique with $x=2\pi F h/H^2$.
In Fig.~\ref{Dingle} we have constructed a Dingle plot by displaying $1/H$ vs 
$\ln\left[AH^{1/2}\sinh\left(\lambda m_{\rm c}^*T/H\right)/J_2(x)\right]$ for the main dHvA branches $\alpha$, $\alpha^\prime$, and $\beta$ at 35~mK for $H\parallel$~$\langle100\rangle$ data.  
From the slope of the Dingle plot, we derived $T_{\rm D}$, which gives an average scattering lifetime $\tau = \hbar/(2\pi k_{\rm B}T_{\rm D})$. 
Knowing $\tau$ and $k_{\rm F}$, the mean free path of electrons can be estimated by $l=\hbar k_{\rm F}\tau /m_{\rm c}^*$.  
We deduce Dingle temperatures of 103, 101, and 102~mK
corresponding to mean free paths of 1903, 1904,and 1950~$\AA$ 
for the $\alpha$, $\alpha^\prime$, and $\beta$ branches, respectively. 
The low Dingle temperature and large mean free path indicate that U$_3$Ni$_3$Sn$_4$ is a clean system with minimal atomic disorder, impurity concentration, or crystal defects. 
%*********************FIGURE 7 **********************************
%****************************************************************
%

Now we evaluate the splitting energy, which corresponds to the magnitude of the antisymmetric spin-orbit interaction, from the simple relation~\cite{Onuki,Mineev,Terashima}
\begin{equation}
\Delta \varepsilon = \frac{\hbar e}{m_{\rm c}^\ast c}\Delta F, 
\end{equation}
where $\Delta F$ is the frequency difference between two splitting branches.
We obtain $\Delta\varepsilon = 24\,{\rm K}$ for branches $\alpha$, $\alpha^\prime$ for $H\parallel \langle 100\rangle$
and $\Delta\varepsilon = 23\,{\rm K}$ for branches $\delta$, $\delta^\prime$ for $H\parallel \langle 110\rangle$.

Here we used the cyclotron effective mass $m_{\rm c}^\ast$ to deduce the splitting energy. 
In the band structure calculations, the band mass is obtained from $m_{\rm b} = \hbar^2/ 2\pi (\Delta S/\Delta \varepsilon)$,
where $S$ is the cross-sectional area of the Fermi surface.
Thus, the splitting energy can be derived using the band mass instead of the cyclotron effective mass.
In this case, we obtained $\Delta\varepsilon = 100$ and $68\,{\rm K}$ 
for branches $\alpha$, $\alpha^\prime$ ($H\parallel \langle 100\rangle$) 
and branches $\delta$, $\delta^\prime$ ($H\parallel \langle 110\rangle$), respectively.

These values are small compared with those for other materials without inversion symmetry.
In VSi$_2$, NbSi$_2$ and TaSi$_2$ which have a chiral structure, the splitting energy has been estimated to be 
$20$--$100$, $200$--$300$, and $500$-$600\,{\rm K}$, respectively from experiments.
The splitting energy is dependent on the $d$-electron, and the $5d$-electrons in TaSi$_2$ produce the largest splitting energy.
A similar trend also occurs in LaTGe$_3$ (T=Co, Rh, Ir).
Naively thinking, it can be inferred that U$_3$Ni$_3$Sn$_4$ has a large splitting energy, 
because of the $5f$ and $6d$-electrons in the U atom.
However, the present results for U$_3$Ni$_3$Sn$_4$ imply that
the splitting energy is strongly reduced by the large cyclotron mass or the band mass.

Note that we assumed the splitting bands are paired, such as 
$\alpha$ and $\alpha^\prime$, 
$\delta$ and $\delta^\prime$, for simplicity.
However, the adjoining bands are not simply paired, as shown in Fig.~\ref{fig:bandW},
because the interband distances are not sufficiently larger than the spin-splitting. 
Thus, it is not simple to derive the splitting energy from the experiment.

\section{Summary}
We succeeded in growing high quality single crystals of U$_3$Ni$_3$Sn$_4$ without inversion symmetry in the crystal structure.
The resistivity, specific heat, and magnetic susceptibility revealed that U$_3$Ni$_3$Sn$_4$ is a paramagnetic heavy-fermion compound in the proximity of the quantum critical point.
The dHvA experiments showed that the Fermi surfaces consist of closed ones,
and the angular dependence of the dHvA frequency is in reasonably good agreement with the LDA band calculations based on the $5f$-itinerant model.
To our knowledge, this is the first time that
the splitting of Fermi surfaces due to the lack of inversion symmetry has been clearly detected in $5f$ electron systems.
The large cyclotron effective mass of up to $35\,m_0$ indicates the heavy electronic state in U$_3$Ni$_3$Sn$_4$.
The small splitting energy is probably related to the heavy quasi-particles in U$_3$Ni$_3$Sn$_4$.

\section*{Acknowledgements}
We thank Y. \={O}nuki, A. de Visser, N. Kimura, and V. P. Mineev for fruitful discussions.
This work was supported by KAKENHI (JP15H05882, JP15H05884, JP15K21732, JP15H05745, JP16H04006, JP15K05156, JP16K17733, JP17K14328).

\end{document}